\documentclass[letterpaper]{aa}
\usepackage{graphicx}
\begin{document}
\title{The diameters of $\alpha$~Centauri A and B}
\subtitle{A comparison of the asteroseismic and VINCI/VLTI views}
\author{P. Kervella \inst{1},
	F. Th\'evenin\inst{2},
	D. S\'egransan\inst{3},
	G. Berthomieu\inst{2},
	B. Lopez \inst{4},
	P. Morel \inst{2}
	\and
	J. Provost \inst{2}
          }
   \offprints{P. Kervella}
   \institute{European Southern Observatory,
              Alonso de Cordova 3107, Casilla 19001, Santiago 19, Chile\\
              \email{Pierre.Kervella@eso.org}
   \and
   D\'epartement Cassini, UMR CNRS 6529, Observatoire de la C\^ote 
   d'Azur, BP 4229, 06304 Nice Cedex 4, France
   \and
   Observatoire de Gen\`eve, CH-1290 Sauverny, Switzerland
   \and
   D\'epartement Fresnel, UMR CNRS 6528, Observatoire de la C\^ote d'Azur,
   BP 4229, 06304 Nice Cedex 4, France
}

\titlerunning{ The diameters of $\alpha$\,Cen\,A \& B}
\authorrunning{Kervella et al.}
\mail{Pierre.Kervella@eso.org}
\date{Received date; accepted date}
\abstract{
We compare the first direct angular diameter measurements obtained on our
closest stellar neighbour, $\alpha$\,Centauri to recent model diameters
constrained by asteroseismic observations.
Using the VINCI instrument installed at ESO's VLT Interferometer (VLTI),
the angular diameters of the two main components of the
system, $\alpha$\,Cen\,A and B, were measured with a relative precision
of 0.2\% and 0.6\% respectively.
Particular care has been taken in the calibration of these measurements,
considering that VINCI is estimating the fringe visibility using a broadband K filter.
We obtain uniform disk angular diameters for $\alpha$\,Cen\,A and B of
$\rm \theta_{\rm UD}[A] = 8.314 \pm 0.016$~mas
and $\rm \theta_{\rm UD}[B] = 5.856 \pm 0.027$~mas,
and limb darkened angular diameters of
$\rm \theta_{\rm LD}[A] = 8.511 \pm 0.020$~mas
and $\rm \theta_{\rm LD}[B] = 6.001 \pm 0.034$~mas.
Combining these values with the parallax from S\"oderhjelm~(\cite{soderhjelm}),
we derive linear diameters of
$D[A] = 1.224 \pm 0.003\ D_{\odot}$ and $D[B] = 0.863 \pm 0.005\ D_{\odot}$.
These values are compatible with the masses published by Th\'evenin et al.~(\cite{thevenin02}) for
both stars.
}

\maketitle
%
\section{Introduction}\label{sec:int}
The $\alpha$\,Centauri triple star system is our closest stellar neighbour. The main
components are G2V and K1V solar-like stars, while the third member is the red
dwarf {\it Proxima} (M5.5V).
$\alpha$\,Cen\,A (\object{HD 128620}) and B (\object{HD 128621})
offer the unique possibility to study the stellar physics at play
in conditions just slightly different from the solar ones. Their masses bracket nicely the
Sun's value, while they are slightly older.
In spite of their high interest, proximity and brightness,
the two main components have never been resolved by
long baseline stellar interferometry, due to their particularly southern position in the sky.
We report in this paper the first direct measurement of their angular diameters.
As a remark, the angular diameter of {\it Proxima} has also been measured recently 
for the first time ($\theta_{\rm LD}=1.02 \pm 0.08$~mas) using two 8-meters
Unit Telescopes and the VINCI instrument (S\'egransan et al.~\cite{s03}).

More than fourty years after the discovery of the solar seismic frequencies
by Leighton~(\cite{leighton60}), and Evans \& Michard~(\cite{em62}),
solar-like $p$ oscillations have been identified on $\alpha$\,Cen\,A \& B by 
Bouchy \& Carrier~(\cite{bc01}, \cite{bc02}) with the CORALIE fiber-fed spectrograph.
Today, asteroseismic frequencies have been detected in several additional stars.
All these observations provide constraints, on one hand on stellar interior studies,
and on the other hand on macroscopic stellar parameters like mass and radius.
Several binary systems like $\alpha$\,Cen (see Morel et al. \cite{m01} for references)
have been calibrated using spectro-photometry constraints.
Recently, $\alpha$\,Cen\,A has been
calibrated using photometry, astrometry, spectroscopy and asteroseismic
frequencies (Th\'evenin et al.~\cite{thevenin02}).
These authors derived the age of the couple, the initial helium abundance $\rm Y_i$,
and the radii of both stars. This calibration was based on stellar evolution
models computed using the CESAM code (Morel \cite{m97}).
One of the main results of this calibration was to constrain the
masses of both stars, and in particular the mass of B.
It had to be diminished by 3\%, compared to the mass proposed by
Pourbaix et al.~(\cite{pnm02}), leading to a smaller diameter of the star B.
The high precision interferometric measurements of the angular diameters of
$\alpha$\,Cen\,A and B with VINCI/VLTI are a direct test of these refined
models.

\section{Depscription of the instrument}
\subsection{The VLT Interferometer and VINCI}
The European Southern Observatory's Very Large Telescope Interferometer
(Glindemann et al. \cite{glindemann}) is operated  on top of the Cerro Paranal,
in Northern Chile, since March 2001. In its current state of completion, the light
coming from two telescopes can be combined coherently in VINCI, the VLT
Interferometer Commissioning Instrument
(Kervella et al.~\cite{k00}, Kervella et al.~\cite{kervella03a}), or in the
mid-infrared instrument MIDI (Leinert et al.~\cite{leinert00}).
A three ways beam combiner, AMBER (Petrov et al.~\cite{petrov00}), will soon be installed
in addition to these instruments.
VINCI uses in general a regular $K$ band filter ($\lambda=2.0-2.4\ \mu$m),
as this was the case for our $\alpha$\,Cen observations, but can also be operated
in the $H$ band ($\lambda=1.4-1.8\ \mu$m) using an integrated optics beam combiner
(Berger et al. \cite{berger01}).
The $K$ band setup effective wavelength changes slightly,
depending on the spectral type of the observed target, between 2.174 and 2.184~$\mu$m
for  3000 $\le$ T$_{\rm eff} \le$ 100000~K.
For $\alpha$\,Cen A and B, $\lambda_{\rm eff} = 2.178 \pm 0.003\ \mu$m (see
Sect.~\ref{ins_transmission} for details).

\subsection{Interferometer configuration}

We used as primary light collectors the two 0.35m Test Siderostats
of the VLTI. After being delayed by the VLTI optical delay lines, the stellar light was
recombined in the interferometric laboratory using the VINCI instrument.
A large number of baselines are accessible on the Cerro Paranal summit.
Two of them were used for this study: E0-G0 and E0-G1,
respectively of 16 and 66 meters ground length. 
The 16m baseline observations were obtained
during the early commissioning phase, from two days to a few weeks
after the first fringes in March 2001. At the time, the
effective aperture of the siderostats was limited to 0.10m
due to the unavailability of optical beam
compression devices. 
Later in 2001, their installation allowed to recover the full 0.35m
primary mirror aperture of the siderostats, and all the 66m baseline
observations reported here were done with the full mirror.
The shorter 16m baseline is useful in the case of $\alpha$\,Cen\,A to determine
unequivocally the position of the 66m measurements on the visibility curve, 
but does not bring a significant contribution to the final angular diameter
precision (see Sect.~\ref{angdiams}). 

\subsection{Visibility calibration}
During observations, the interferometric efficiency (visibility produced by the
system when observing a point source)
varies slowly over a timescale of hours. This means that the scientific
target observations have to be calibrated periodically
using observations of a star with a known angular diameter.
The data reduction software of VINCI yields accurate estimates of the squared
modulus of the coherence factor $\mu^{2}$, which is linked to the object visibility
$V$ by the relationship
\begin{equation}
V^{2} = \frac{\mu^2}{T^2}
\end{equation}
where $T$ is the interferometric efficiency.
$T$ is estimated by bracketing the science target with observations of calibrator
stars whose $V$ is supposed to be known a priori. The precision of our knowledge
of the calibrator's angular diameter is therefore decisive for the final quality of the
calibrated visibility value. For our observations, we have applied a constant
transfer function $T^{2}$ between calibrator stars. This assumption
has been validated during routine VLTI observations. A detailed description of
the calibration observations can be found in Sect.~\ref{calib_observ}.

\section{Data processing}\label{fringes_calib}
\subsection{Data processing algorithm}
We used a customized version of the standard VINCI data
reduction pipeline (Kervella et al. \cite{k03}), whose general principle
is based on the original FLUOR algorithm (Coud\'e du Foresto et al. \cite{cdf97}).
The two calibrated output interferograms are subtracted to remove residual
photometric fluctuations. We implemented in this code
a time-frequency analysis (S\'egransan et al. \cite{s99})
based on the continuous wavelet transform (Farge \cite{farge92}).
Instead of the projection of the signal onto a sine wave of the Fourier transform, the wavelet
transform decomposes it onto a function, i.e. the wavelet, that is localised in both
time and frequency. We used as a basis the Morlet wavelet, a gaussian envelope
multiplied by a sinus wave. With the proper choice of the number of oscillations
inside the gaussian envelope, the Morlet wavelet closely matches a VINCI
interferogram. It is therefore very efficient at localizing the signal in both time and frequency.
%
\subsection{Data quality control}
In spite of the relatively high modulation frequency of the fringes (296 Hz for
the 66m baseline measurements), a fraction of the recorded interferograms present
a differential piston signature between the two apertures.
This is due to the relatively low coherence time observed at Paranal (1-4 ms at $\lambda$=500 nm).
These interferograms are rejected in the VINCI data processing by comparing the
frequency of the measured fringe peak with the expected frequency from the K band
filter of VINCI. If the measured fringe frequency is different by more than 20\%
from the expected frequency, the interferogram is ignored.
The fringe packet extensions in the time and frequency domains
are also used for the selection.
This process allows to keep only the best quality interferograms and
reduces the final dispersion of the visibilities.
Finally, we rejected the observations that presented an abnormally
low photometric signal to noise ratio, that is a typical symptom of an inaccuracy
in the pointing of the siderostats.

The total numbers of selected and processed interferograms are
7854 on $\alpha$\,Cen\,A (2427 on the 66m baseline and 5427 on the 16m baseline)
and 1833 on $\alpha$\,Cen\,B (66m baseline only).
For the calibration of the $\alpha$\,Cen
observations, we processed 2998 interferograms of $\theta$\,Cen.
Several calibrators (including $\theta$\,Cen) were used for the 16m baseline
measurements of $\alpha$\,Cen\,A, for a total of 8059 processed interferograms.
The separate measurement of $\theta$\,Cen was achieved using 1750 interferograms
of this star and 789 interferograms of the secondary calibrator 58\,Hya.

%
\begin{figure}
\centering
\includegraphics[bb=0 0 360 288, width=8.5cm]{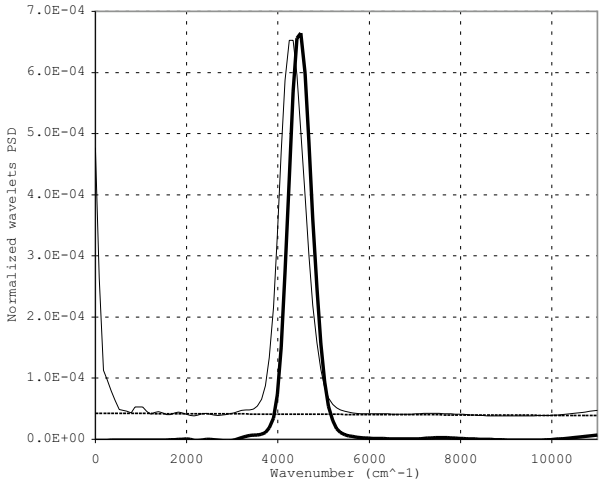}
\caption{Wavelets power spectral density (PSD) of a processed series of 418 interferograms
obtained on $\alpha$\,Cen\,A, before (thin line) and after (thick line) recentering of the
fringe peak and subtraction of the background noise (dashed line).
}
\label{Processed_peak_A}
\end{figure}
After the processing of a series of interferograms (100 to 500 scans),
the mean squared coherence factor is derived
from the average wavelets power spectral density of the selected interferograms.
Fig.~\ref{Processed_peak_A} shows the average wavelets
power spectral density (PSD) of 418 processed interferograms,
summed over the time extent of the fringe packet
to obtain a one dimensional vector. In spite of the very low visibility of the
fringes of $\alpha$\,Cen\,A on the 66m baseline ($V^2 \simeq 1\%$), the fringe peak
is well defined.

The noise background (residual detector and photon shot noise) is
estimated directly from the higher and lower frequencies of the average PSD of the
interferogram,  and then subtracted. As shown on Fig.~\ref{Processed_peak_A}, the
subtraction is very efficient and gives a clean final PSD. The individual
interferogram PSDs are summed after recentering of the fringe peak maximum,
to reduce the power spread due to piston effect. This avoids that energy
is lost in the integration process and allows a more precise estimation
of the background level.
We have chosen not to use the background removal method described
by Perrin (\cite{perrin03}), as we are simultaneously removing both the photon shot
noise and detector noise contributions.

On Fig.~\ref{Processed_peak_A}, the recentered and background corrected
fringe peak is shifted slightly towards higher wavenumber values due to
the variation of $\alpha$\,Cen\,A visibility over the $K$ band.
For simplicity reasons, the data reduction software assumes a fixed
wavelength of 2.195~$\mu$m of the fringe peak maximum for the recentering
process for all stars, but the exact target value has no effect on the final visibility.

\subsection{Instrumental transmission}\label{ins_transmission}
\subsubsection{Transmission model}
When using VINCI, the observations are carried out using a full $K$ band filter, accepting the
star light from 2.0 to 2.4 $\mu$m. In order to obtain a precise fit of the calibrated
visibilities measured on sky, we computed the transmission of the interferometer
taking into account the atmospheric transmission (Lord~\cite{lord92}), the fluoride
glass optical fibers, the $K$ band filter and the quantum efficiency of the HAWAII detector.
This gave us a first approximation of the transmission of the
interferometer $F_0(\lambda)$.
\subsubsection{On-sky calibration}
The instrumental uncertainties led us to compare directly this theoretical
VINCI/VLTI transmission model to the real transmission of the system on sky.
This has been achieved through the precise estimation of the effective
wavenumber of a series of bright stars observations obtained with VINCI
and two 8 meters Unit Telescopes (Table \ref{Transmission}).
A multiplicative slope $\gamma$ (expressed in $\mu$m$^{-1}$) is superimposed
to the theoretical transmission $F_0(\lambda)$ in order to match the
observed average position of the PSD fringe peak.
It is the only variable adjusted to match the observations.
The photometric signal to noise ratio of the UTs observations
being very high, we obtain a good precision on the average fringe
peak frequency and thus on the estimation of $\gamma$,
as shown in Table \ref{Transmission}. The total photometric
transmission of the interferometer $F(\lambda)$ is then given by:
\begin{equation}
F(\lambda) = \gamma (\lambda - \lambda_0)\ F_0(\lambda)
\end{equation}
The reference wavelength $\lambda_0$ was set arbitrarily to $1.90\ \mu$m in our
computation, but has no influence on the transmission calibration.
$F(\lambda)$ is not normalized and gives only relative transmission values over
the K band, but the absolute transmission is not needed to derive the model visibilities.
It should be stressed that the sensitivity of the final angular
diameter to $\gamma$ is low, a $\pm 0.08\ \mu$m$^{-1}$ change of this parameter
resulting only in a $\pm 0.010$~mas change on $\theta_{\rm UD}$ for $\alpha$\,Cen\,A, and
$\pm 0.007$~mas for B. These uncertainties were quadratically added
to the final errors of the UD and LD angular diameter values given further
in this paper. As a remark, we can also express this as a $\pm 0.003\ \mu$m uncertainty
on the average value of $\lambda_{\rm eff} = 2.178\ \mu$m that we derive
from $F(\lambda)$ for $\alpha$\,Cen\,A and B.

\begin{table}
\caption[]{
Determination of the transmission correction slope $\gamma$
of the VINCI/VLTI combination as observed on bright stars with
two 8-meters telescopes.}
\label{Transmission}\begin{tabular}{lcc}
\hline 
& Peak position ($\mu$m)& $\gamma$ ($\mu$m$^{-1}$)\\
\hline
$\alpha$~PsA & 2.198 $\pm$ 0.002 & 1.140\\
HR 8685 (1) & 2.190 $\pm$ 0.003 & 0.911\\
HR 8685 (2) & 2.202 $\pm$ 0.008 & 1.119\\
$\gamma^{2}$ Vol & 2.198 $\pm$ 0.007 & 1.076\\
$\epsilon$ Eri & 2.197 $\pm$ 0.010 & 1.065\\
39 Eri & 2.196 $\pm$ 0.010 & 1.076\\
\hline
& Weighted average & 1.076 $\pm$ 0.081\\
\hline
\end{tabular}
\end{table}
\subsubsection{Discussion}
The observations of $\alpha$\,Cen have been obtained with the siderostats, that
have a slightly different optical setup than the UTs.
There are 26 reflections for the UTs configuration in each arm of the
interferometer, compared to 20 for the siderostats. Out of these,
15 mirrors are common between the two configurations. The remaining difference
is therefore between the additional 11 reflections of the UTs and the additional 5
reflexions of the siderostats. Even assuming a very conservative mismatch
of 1\% between the extreme wavelengths reflectivity of each mirror of the UT
train compared to the siderostats, we obtain a relative difference on $\gamma$
of only 3\% that is significantly less than our quoted statistical uncertainty (7.5\%).
We have therefore considered this diffference negligible in our study.

A possible reason for the observed wavelength drift is the aging of the 20 mirror coatings
necessary to bring the star light into the VINCI instrument (for each of the two beams).
This process may have affected differentially the reflectivity of one end
of the $K$ band compared to the other. A difference in reflectivity of only 1\%
between the two extreme wavelengths will result in an 18\% difference on the final
transmission, after 20 mirrors (siderostats configuration).
Also, the transmission curves provided by the manufacturer of the fiber
optics used in VINCI do not have a sufficient precision to constrain
accurately the instrument transmission model, and an error of several tens
of percent is not to be excluded.
To a lesser extent, the engineering grade HAWAII infrared array used in
VINCI may have a quantum efficiency curve
differing from the science grade versions by several percent. 
Finally, the MONA triple coupler used to recombine the light has also shown
some birefringence during laboratory tests. This effect could result in a shift
of the effective observation wavelength.

To secure the internal wavelength calibration of VINCI itself, crucial for the
accuracy of the estimation of $\gamma$, we have obtained laboratory fringes with
a K band laser ($\lambda = 2.304\ \mu$m). This gave us a precise wavelength reference
to verify the scanning speed and the camera acquisition frequency.
\subsubsection{Source spectrum}
\begin{figure}
   \centering
   \includegraphics[bb=0 0 360 288, width=8.5cm]{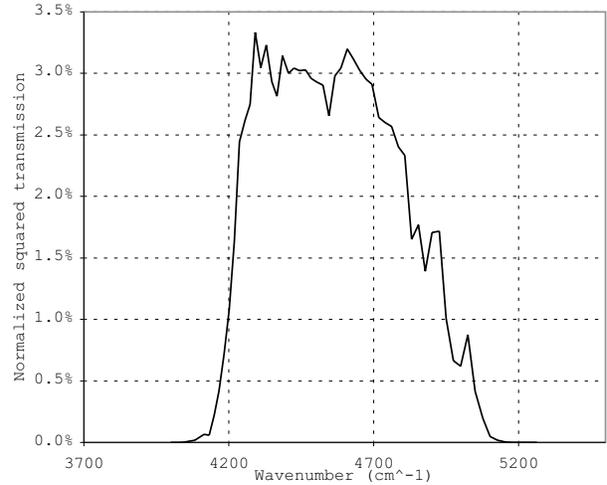}
      \caption{Model PSD of $\alpha$\,Cen\,A fringes for a zero baseline, 
including the spectrum of the star, atmosphere, fluoride glass fibers, K band filter, detector
quantum efficiency and the correction function $F_1(\lambda)$ (see Sect.~\ref{ins_transmission}).
A total precipitable water vapor of 3.0 mm (median for Paranal)
is assumed for this plot. The curve for $\alpha$~Cen B is almost identical.}
         \label{T2_raw}
   \end{figure}

In addition to the constant term of the instrumental transmission, the shape of the source
spectrum for each star was taken into account using its effective temperature.
We computed synthetic spectra for $\alpha$\, Cen~A and B using Kurucz models,
but considering the absence of any large absorption feature in the $K$ band,
we did not include spectral features in our final transmission model.
The simulated spectrum of $\alpha$\,Cen~A fringes for a zero baseline
is shown on Fig.~\ref{T2_raw}.

\subsection{Bandwidth smearing}\label{bandwidth_smearing}
\begin{figure}
   \centering
   \includegraphics[bb=0 0 360 288, width=8.5cm]{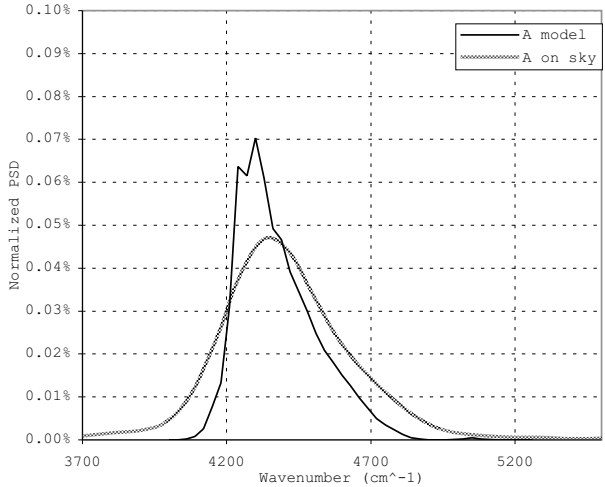}
      \caption{Model (black line) and observed (grey line) PSDs of $\alpha$\,Cen\,A
fringes on the 66m baseline (61m projected).
The visibility loss for larger wavenumbers is clearly visible.
The observed Fourier PSD, smoothed by the differential piston, shows the expected asymmetry.
The on-sky and model vertical scales are arbitrary.}
         \label{Spectrum_A}
   \end{figure}

\begin{figure}
   \centering
   \includegraphics[bb=0 0 360 288, width=8.5cm]{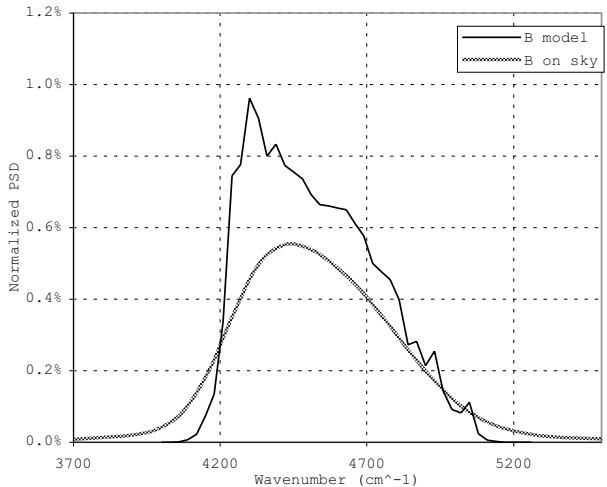}
      \caption{Model (black line) and observed (grey line) PSDs of
$\alpha$\,Cen\,B fringes on the E0-G1 baseline.
$\alpha$\,Cen\,B being significantly less resolved than A, the squared
visibility is more uniform over the K band than
for $\alpha$~Cen A (Fig.~\ref{Spectrum_A}).
The asymmetry of the power spectrum is therefore smaller, though still present.
The on-sky and model vertical scales are arbitrary.}
         \label{Spectrum_B}
   \end{figure}

An important effect of the relatively large spectral bandwidth of the VINCI
filter is that several spatial frequencies are observed simultaneously
by the interferometer. This effect is called {\it bandwidth smearing}.
In the case of $\alpha$\,Cen\,A, it is particularly strong as the visibilities
are close to the first minimum of the visibility function, and this effect cannot be neglected.
With a 60m projected baseline, the short wavelength edge of the K band
($\lambda \simeq 2.0 \mu$m) is already at the null of visibility, while
the $V^2$ for the long wavelength edge ($\lambda \simeq 2.4 \mu$m) is still above 1\%.
The photons at the null of visibility have interfered destructively.
Therefore, the fringe peak becomes very asymmetric
in the PSD of the interferograms.
As shown on Fig.~\ref{Spectrum_A} and \ref{Spectrum_B}, 
the observed and model PSDs agree well in general shape.
The on-sky power spectrum is blurred by the differential piston and therefore
appears "smoothed", but the characteristic asymmetry for low visibilities
is clearly visible.
\subsection{Baseline smearing}
When the aperture of the light collectors is a significant fraction of the baseline,
an effect similar to the bandwidth smearing appears on the visibility
measurements. It comes from the fact that the baselines defined between
different parts of the two primary mirrors cover a non-zero range of lengths
and orientations. Therefore, several spatial frequencies are measured
simultaneously by the beam combiner.
In the case of the E0-G0 baseline (16m) observations of $\alpha$\,Cen\,A, the effective aperture
was 0.10m, and therefore the ratio of the primary mirror diameter to the baseline
was only $D/B \approx 0.6$\%.
For the E0-G1 baseline (66m), this ratio is similar due to the larger 0.35m apertures.
Even in the difficult case of the $\alpha$\,Cen\,A observations, this effect accounts at
most for a relative shift of the visibility of 0.1\%, to be compared to our relative
systematic calibration error of 1.5\%. In the case of $\alpha$\,Cen\,B, we expect
at most a 0.05\% shift, for a relative systematic calibration error also of 1.5\%.
We have therefore neglected this effect in the rest of our study.


%
%
%

\section{Calibration observations}\label{calib_observ}
%
The calibration of the interferometric efficiency (IE) of the interferometer is a critical
step of the observations. We present in Table~\ref{Cal_alfcen} the measurements that we obtained
on the calibrators and the corresponding values of the IE for the three nights of observations
of the $\alpha$\,Cen pair (JD = 2452462-70) on the E0-G1 baseline,
and the separate night used to measure $\theta$\,Cen (JD = 2452452).

The primary calibrator $\theta$\,Cen is located at a distance of 24 degrees
from the $\alpha$\,Cen pair, mostly in declination, while only 9 degrees
separate $\theta$\,Cen and the secondary calibrator 58\,Hya.
During the observations, the largest difference between the altitudes of
$\theta$\,Cen and $\alpha$\,Cen happens at the crossing of the meridian, and
is approximately 24 degrees (respective altitudes of about
55 and 80 degrees at Paranal). The airmasses of the two stars at meridian crossing
are 1.25 and 1.03 respectively for $\alpha$\,Cen and $\theta$\,Cen. The difference
is even smaller in the case of 58\,Hya and $\theta$\,Cen.
As we obtained the E0-G1 baseline observations close to the meridian crossing,
we do not expect any significant variation of IE due to the difference of
airmass between the calibrators and the scientific targets.

\begin{table*}
\caption{Calibration observations of $\theta$\,Cen and 58\,Hya.
The expected visibilities given in this table include the bandwidth smearing effect.
The resulting interferometric efficiencies assumed for the calibration of the
$\alpha$\,Cen and $\theta$\,Cen observations are given in bold characters, with the corresponding
statistical and systematic error bars for each observing session.
The calibrated visibilities can be found in Table~\ref{Vis_tetcen} for $\theta$\,Cen, and
in Tables~\ref{table_alfcenA} and \ref{table_alfcenB} for $\alpha$\,Cen\,A and $\alpha$\,Cen\,B.}
\label{Cal_alfcen}
\begin{tabular}{lccccccl}
\hline
JD & Scans & B (m) & Azim. & $\mu^{2}$ (\%) & Expected $V^2$ (\%) & IE ($\mu^2 / V^2$, \%) & Target(s)\\ 
- 2450000& & & (N=0) & $\pm$ stat. err. & $\pm$ syst. err. & $\pm$ stat. $\pm$ syst. & \\
\hline
\noalign{\smallskip}
 & & & & & & {\bf 52.08 $\pm$ 0.69 $\pm$ 0.63} & $\theta$\,Cen \\
2452.63024 & 45 & 65.2509 & 161.90 & 30.72 $\pm$ 0.75 & 58.90 $\pm$ 0.71 & 52.17 $\pm$ 1.27 $\pm$ 0.63 & 58\,Hya\\
2452.63324 & 166 & 65.2011 & 162.55 & 31.56 $\pm$ 0.74 & 58.95 $\pm$ 0.71 & 53.54 $\pm$ 1.25 $\pm$ 0.64 & 58\,Hya\\
2452.64109 & 269 & 65.0719 & 164.26 & 30.66 $\pm$ 0.31 & 59.08 $\pm$ 0.71 & 51.91 $\pm$ 0.52 $\pm$ 0.62 & 58\,Hya\\
2452.64527 & 256 & 65.0047 & 165.18 & 30.73 $\pm$ 0.37 & 59.15 $\pm$ 0.71 & 51.95 $\pm$ 0.62 $\pm$ 0.62 & 58\,Hya\\
 & & & & & & {\bf 52.13 $\pm$ 0.69 $\pm$ 0.63} & $\theta$\,Cen \\
2452.67563 & 53 & 64.5970 & 174.54 & 31.42 $\pm$ 0.77 & 59.56 $\pm$ 0.70 & 52.77 $\pm$ 1.29 $\pm$ 0.62 & 58\,Hya\\
\hline
\noalign{\smallskip}
2462.53152 & 389 & 65.7824 & 153.61 & 8.58 $\pm$ 0.17 & 18.13 $\pm$ 0.28 & 47.33 $\pm$ 0.93 $\pm$ 0.72 & $\theta$\,Cen\\
2462.60245 & 88 & 65.8892 & 168.69 & 8.63 $\pm$ 0.15 & 18.01 $\pm$ 0.27 & 47.94 $\pm$ 0.86 $\pm$ 0.73 & $\theta$\,Cen\\
 & & & & & & {\bf 47.51 $\pm$ 0.49 $\pm$ 0.72} & $\alpha$\,Cen\,A, $\alpha$\,Cen\,B \\
2462.60554 & 283 & 65.8801 & 169.39 & 8.46 $\pm$ 0.21 & 18.02 $\pm$ 0.27 & 46.97 $\pm$ 1.18 $\pm$ 0.72 & $\theta$\,Cen\\
\hline
\noalign{\smallskip}
2465.56068 & 94 & 65.9551 & 161.27 & 8.67 $\pm$ 0.14 & 17.93 $\pm$ 0.27 & 48.33 $\pm$ 0.80 $\pm$ 0.74 & $\theta$\,Cen\\
2465.56377 & 355 & 65.9541 & 161.93 & 8.58 $\pm$ 0.24 & 17.93 $\pm$ 0.27 & 47.87 $\pm$ 1.31 $\pm$ 0.73 & $\theta$\,Cen\\
2465.57390 & 241 & 65.9413 & 164.14 & 8.98 $\pm$ 0.50 & 17.95 $\pm$ 0.27 & 50.06 $\pm$ 2.80 $\pm$ 0.77 & $\theta$\,Cen\\
2465.57838 & 317 & 65.9320 & 165.13 & 8.85 $\pm$ 0.29 & 17.96 $\pm$ 0.27 & 49.31 $\pm$ 1.64 $\pm$ 0.75 & $\theta$\,Cen\\
 & & & & & & {\bf 48.64 $\pm$ 1.50 $\pm$ 0.76} & $\alpha$\,Cen\,A, $\alpha$\,Cen\,B \\
2465.65159 & 69 & 65.8006 & 1.99 & 9.35 $\pm$ 0.44 & 18.11 $\pm$ 0.28 & 51.66 $\pm$ 2.43 $\pm$ 0.79 & $\theta$\,Cen\\
\hline
\noalign{\smallskip}
2470.56229 & 87 & 65.9373 & 164.59 & 9.01 $\pm$ 0.13 & 17.95 $\pm$ 0.27 & 50.19 $\pm$ 0.73 $\pm$ 0.77 & $\theta$\,Cen\\
2470.56609 & 386 & 65.9289 & 165.43 & 9.00 $\pm$ 0.05 & 17.96 $\pm$ 0.27 & 50.09 $\pm$ 0.30 $\pm$ 0.77 & $\theta$\,Cen\\
2470.57010 & 341 & 65.9188 & 166.32 & 9.06 $\pm$ 0.15 & 17.97 $\pm$ 0.27 & 50.44 $\pm$ 0.83 $\pm$ 0.77 & $\theta$\,Cen\\
2470.57433 & 348 & 65.9073 & 167.27 & 8.73 $\pm$ 0.14 & 17.99 $\pm$ 0.27 & 48.53 $\pm$ 0.77 $\pm$ 0.74 & $\theta$\,Cen\\
 & & & & & & {\bf 49.97 $\pm$ 0.87 $\pm$ 0.76} & $\alpha$\,Cen\,A, $\alpha$\,Cen\,B \\
\hline
\end{tabular}
\end{table*}

\section{The primary calibrator $\theta$\,Centauri}
The most important calibrator for the 66 meters baseline measurements is the
giant star $\theta$\,Cen (K0III).
This calibrator was chosen for its stability and brightness
in the list of standard stars compiled by Cohen et al.~(\cite{cohen99})
and verified by Bord\'e et al.~(\cite{borde}).
This choice is critical in the sense that any
departure of the true visibility of the calibrator from the assumed model
will contaminate the calibrated visibility of the scientific target. 
This is the reason why one should avoid to use as calibrators
pulsating variables (such as many M type giants, Cepheids,..),
double or multiple stars, magnetically
active objects (photospheric spots) or fast rotators
(ellipticity of the star disk).
The properties of all the stars listed in the Cohen et al.~(\cite{cohen99})
catalogue have been checked carefully
and their diameters are believed to be constant
to a very good accuracy. In addition, $\theta$\,Cen is not
classified as double, variable or active in any catalogue, and is a slow rotator
($V \sin i = 1.2\ {\rm km.s}^{-1}$, Glebocki et al.~\cite{glebocki00}).

\begin{figure}
   \centering
   \includegraphics[bb=0 0 360 288, width=8.5cm]{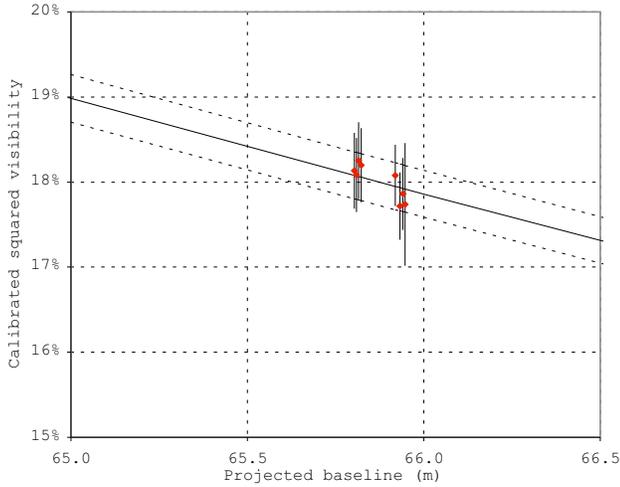}
      \caption{Detail of $\theta$\,Cen squared visibilities and uniform disk model. The continuous line is the UD
diameter fit (5.305 $\pm$ 0.020 mas), and the dotted lines represent the limits of the $\pm 1 \sigma$ error domain.}
         \label{FigTetCen_66m}
   \end{figure}

Unfortunately, the typical 1\% precision of the Cohen et al.~(\cite{cohen99})
catalogue on the angular diameters,
though very good in itself,  is not sufficient due to the large size of this
star and the correspondingly low visibility on the 66 meters baseline.
After the first processing of our $\alpha$\,Cen data, it appeared that the
error bars on the final angular diameters were dominated by
the systematic uncertainty on the angular size of $\theta$\,Cen.
Therefore, we reduced additional archived data obtained
on $\theta$\,Cen on a separate night, using the secondary
calibrator 58\,Hya and the 66 meters baseline.
58\,Hya has a much smaller angular diameter than $\theta$\,Cen and therefore
provides a precise calibration of the interferometric efficiency.
The calibrated squared visibility values obtained on $\theta$\,Cen are
listed in Table~\ref{Vis_tetcen}, and the angular diameter fit is shown
on Fig.~\ref{FigTetCen_66m}.
The parameters for both stars and the measured uniform disk (UD) angular
diameter of $\theta$\,Cen are presented in Table~\ref{tetcen}.
The VINCI/VLTI angular diameter found for this star agrees very well with the Cohen
et al. (\cite{cohen99}) value, while reducing significantly its uncertainty.

\begin{table}\caption{$\theta$\,Cen squared visibilities.}
\label{Vis_tetcen}
\begin{tabular}{cccc}
\hline
JD & B (m) & Azim. & $V^{2}$ (\%)\\ 
- 2450000& & (N=0) & $\pm$ stat. $\pm$ syst. \\
\hline
\noalign{\smallskip}
E0-G1\\
2452.60644 & 65.9464 & 163.49 & 17.74 $\pm$ 0.69 $\pm$ 0.21\\
2452.60943 & 65.9413 & 164.15 & 17.86 $\pm$ 0.36 $\pm$ 0.22\\
2452.61396 & 65.9318 & 165.15 & 17.72 $\pm$ 0.34 $\pm$ 0.21\\
2452.61906 & 65.9193 & 166.28 & 18.08 $\pm$ 0.29 $\pm$ 0.22\\
2452.65518 & 65.8220 & 174.54 & 18.20 $\pm$ 0.38 $\pm$ 0.22\\
2452.65855 & 65.8156 & 175.32 & 18.25 $\pm$ 0.40 $\pm$ 0.22\\
2452.66275 & 65.8088 & 176.30 & 18.08 $\pm$ 0.38 $\pm$ 0.22\\
2452.66685 & 65.8037 & 177.26 & 18.13 $\pm$ 0.39 $\pm$ 0.22\\
\hline
\end{tabular}\end{table}

   \begin{table}
      \caption[]{Parameters of the primary ($\theta$~Cen) and secondary
(58~Hya) calibrators.}
	\label{tetcen}	
\begin{tabular}{lcc}
& $\theta$\,Cen & 58 Hya\\
& \object{HD 123139} & \object{HD 130694}  \\
\hline
\noalign{\smallskip}
$m_\mathrm{V}$ & 2.06 & 4.42\\
$m_\mathrm{K}$ & -0.10 & 1.13\\
Spectral Type & K0IIIb & K4III\\
$\rm T_{\mathrm{eff}}$ (K)$^{\mathrm{a}}$ & 4980 & 4040\\
Measurement $\lambda$ ($\mu$m) & 2.181 & 2.181 \\
$\log g$ $^{\mathrm{a}}$ & 2.75 & 1.85 \\
$\theta_{\rm {LD}}$ (mas)$^{\mathrm{b}}$  & 5.46 $\pm$ 0.058 & 3.22 $\pm$ 0.035\\
$\theta_{\rm {UD}}$ (mas)$^{\mathrm{c}}$ & 5.33 $\pm$ 0.057 & 3.12 $\pm$ 0.034\\
Measured $\theta_{\rm {UD}}$ (mas) & {\bf 5.305} ${\bf \pm}$ {\bf 0.020} &\\
            \noalign{\smallskip}
            \hline
\end{tabular}
\begin{list}{}{}
\item[$^{\mathrm{a}}$] $\rm T_{\mathrm{eff}}$ and $\log g$ 
from Cayrel de Strobel et al. (\cite{cayrel}).
\item[$^{\mathrm{b}}$] Catalogue value from Cohen et al. (\cite{cohen99}).
\item[$^{\mathrm{c}}$] Linear limb darkening from Claret (\cite{claret00}).
\end{list}
\end{table}

\section{Calibrated visibilities}
The list of the observations of $\alpha$~Cen A and B, with the resulting calibrated squared
visibilities, is presented in Tables \ref{table_alfcenA} and \ref{table_alfcenB}.
The azimuth of the projected baseline is counted clockwise (cw) from north,
and corresponds to the baseline orientation as seen from the star.
Two error bars are given for each $V^2$ value:
\begin{itemize}
\item one statistical error bar, computed from the dispersion of the visibility values obtained during the observation,
\item one systematic error bar defined by the uncertainty on the knowledge of the calibrator angular size.
\end{itemize}
While the statistical error can be diminished by repeatedly observing the target, the systematic
error cannot be reduced by averaging measurements obtained using the same calibrator. This
is taken into account in our model fitting by checking that the final uncertainty
of the fit is larger than the systematic errors of each measured visibility value. This
conservative approach ensures that we are not underestimating the systematic
calibration errors.

\begin{table}\caption{$\alpha$\,Cen A squared visibilities, expressed in percents.}
\label{table_alfcenA}
\begin{tabular}{cccc}
\hline
JD & B (m) & Azim. & $V^{2}$ (\%)\\ 
- 2450000& & (N=0) & $\pm$ stat. $\pm$ syst. \\
\hline
\noalign{\smallskip}
E0-G0\\
1988.78108 & 15.9201 & 64.95 & 78.99 $\pm$ 1.48 $\pm$ 2.81\\
1988.78380 & 15.9071 & 65.79 & 79.61 $\pm$ 1.46 $\pm$ 2.83\\
1988.78652 & 15.8930 & 66.62 & 79.82 $\pm$ 1.42 $\pm$ 2.84\\
1988.78901 & 15.8793 & 67.39 & 79.80 $\pm$ 1.47 $\pm$ 2.84\\
1995.76493 & 15.9058 & 65.86 & 80.15 $\pm$ 1.03 $\pm$ 0.66\\
1996.63335 & 15.7916 & 24.31 & 78.66 $\pm$ 1.10 $\pm$ 0.42\\
1996.63970 & 15.8129 & 26.48 & 82.19 $\pm$ 1.12 $\pm$ 0.44\\
1996.64733 & 15.8390 & 29.06 & 80.33 $\pm$ 1.29 $\pm$ 0.43\\
1996.65492 & 15.8650 & 31.61 & 81.49 $\pm$ 1.15 $\pm$ 0.44\\
1996.67842 & 15.9399 & 39.40 & 80.87 $\pm$ 1.10 $\pm$ 0.43\\
1996.68327 & 15.9532 & 40.99 & 79.40 $\pm$ 1.10 $\pm$ 0.43\\
2001.80688 & 15.3644 & 83.76 & 80.71 $\pm$ 1.94 $\pm$ 0.05\\
2001.80954 & 15.3273 & 84.59 & 82.77 $\pm$ 2.69 $\pm$ 0.04\\
2002.70376 & 16.0062 & 52.80 & 78.76 $\pm$ 0.74 $\pm$ 0.05\\
2002.70640 & 16.0057 & 53.63 & 80.01 $\pm$ 1.07 $\pm$ 0.05\\
2003.83537 & 14.8150 & 94.44 & 82.49 $\pm$ 1.10 $\pm$ 0.05\\
2003.83780 & 14.7695 & 95.22 & 82.01 $\pm$ 1.80 $\pm$ 0.04\\
2003.84099 & 14.7088 & 96.25 & 85.30 $\pm$ 1.28 $\pm$ 0.05\\
2003.84356 & 14.6589 & 97.08 & 83.64 $\pm$ 1.77 $\pm$ 0.04\\
\\ E0-G1\\
2462.55258 & 59.2848 & 150.05 & 1.132 $\pm$ 0.051 $\pm$ 0.017\\
2462.55613 & 59.4391 & 150.91 & 1.139 $\pm$ 0.032 $\pm$ 0.017\\
2462.56087 & 59.6365 & 152.05 & 1.099 $\pm$ 0.031 $\pm$ 0.017\\
2462.56493 & 59.7975 & 153.04 & 1.054 $\pm$ 0.029 $\pm$ 0.016\\
2465.61044 & 61.2943 & 166.21 & 0.626 $\pm$ 0.035 $\pm$ 0.010\\
2470.58454 & 61.0497 & 163.19 & 0.758 $\pm$ 0.052 $\pm$ 0.012\\
2470.60337 & 61.4043 & 167.84 & 0.624 $\pm$ 0.023 $\pm$ 0.010\\
2470.60778 & 61.4696 & 168.93 & 0.637 $\pm$ 0.033 $\pm$ 0.010\\
\hline
\end{tabular}\end{table}

\begin{table}\caption{$\alpha$\,Cen B squared visibilities.}
\label{table_alfcenB}
\begin{tabular}{cccc}
\hline
JD & B (m) & Azim. & $V^{2}$ (\%)\\ 
- 2450000& & (N=0) & $\pm$ stat. $\pm$ syst. \\
\hline
\noalign{\smallskip}
E0-G1\\
2462.58356 & 60.4413 & 157.57 & 17.02 $\pm$ 0.36 $\pm$ 0.26\\
2462.58697 & 60.5443 & 158.40 & 17.01 $\pm$ 0.23 $\pm$ 0.26\\
2462.59047 & 60.6453 & 159.26 & 16.80 $\pm$ 0.77 $\pm$ 0.26\\
2462.59490 & 60.7665 & 160.35 & 16.05 $\pm$ 0.68 $\pm$ 0.24\\
2465.62682 & 61.5409 & 170.27 & 16.76 $\pm$ 1.05 $\pm$ 0.26\\
2470.62033 & 61.6208 & 172.05 & 14.94 $\pm$ 0.44 $\pm$ 0.23\\
2470.62342 & 61.6500 & 172.82 & 15.59 $\pm$ 0.42 $\pm$ 0.24\\
2470.62783 & 61.6866 & 173.92 & 16.70 $\pm$ 0.44 $\pm$ 0.25\\
\hline
\end{tabular}\end{table}
%
\section{Angular diameters}\label{angdiams}
\subsection{Uniform disk}
   \begin{figure}
   \centering
   \includegraphics[bb=0 0 360 288, width=8.5cm]{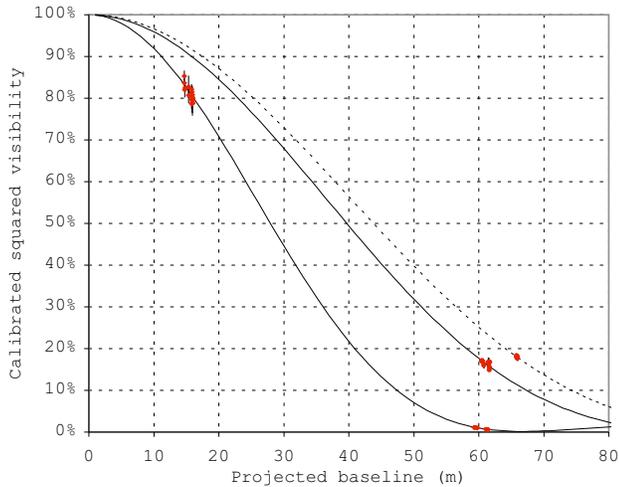}
      \caption{Overview of the $\alpha$\,Cen and $\theta$\,Cen squared visibilities and UD models. From bottom to top:
$\alpha$\,Cen\,A, $\alpha$\,Cen\,B and $\theta$\,Cen (primary calibrator). The angular diameter of $\theta$\,Cen
was measured using 58\,Hya as secondary calibrator.}
         \label{FigAlfCenA_16m}
   \end{figure}

   \begin{figure}
   \centering
   \includegraphics[bb=0 0 360 288, width=8.5cm]{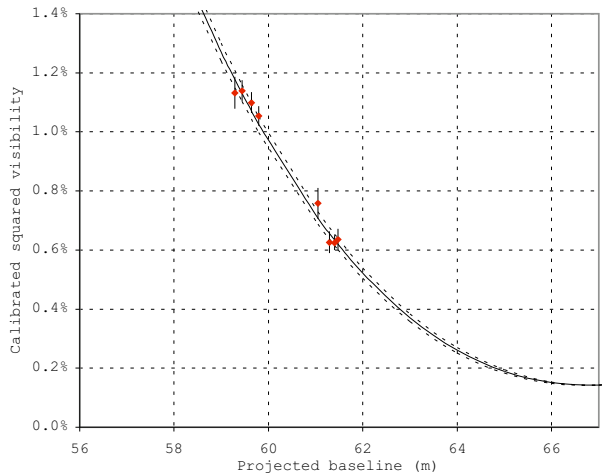}
      \caption{Detail of $\alpha$\,Cen\,A squared visibility. The continuous line is the uniform disk diameter
fit (8.314 $\pm$ 0.016 mas), and the dotted lines represent the limits of the $\pm 1 \sigma$ error domain. The
visibility curve never goes to zero due to the bandwidth smearing effect.}
         \label{FigAlfCenA_66m}
   \end{figure}

   \begin{figure}
   \centering
   \includegraphics[bb=0 0 360 288, width=8.5cm]{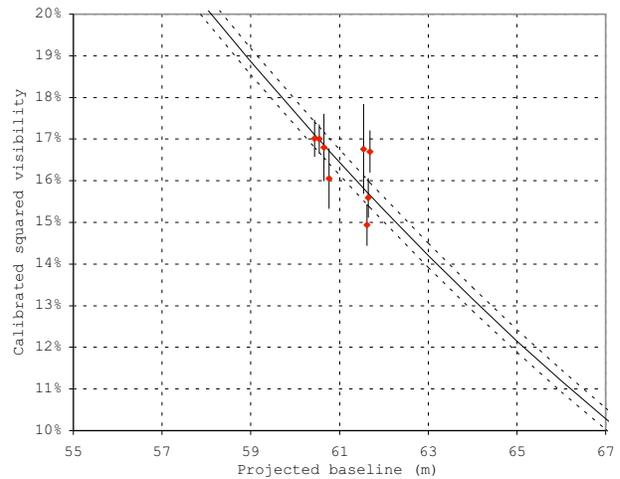}
      \caption{Detail of $\alpha$\,Cen\,B squared visibility. The continuous line is the uniform disk
diameter fit (5.856 $\pm$ 0.027 mas), and the dotted lines represent the limits of the $\pm 1 \sigma$ error domain.}
         \label{FigAlfCenB_66m}
   \end{figure}
Due to the spectrum shape variation with baseline described in Sect.~\ref{bandwidth_smearing},
the classical monochromatic uniform disk (UD) model visibility curve is not applicable
and can lead to very large UD size errors for low visibilities.
We therefore adopted a direct fitting method in the broadband regime.
For this purpose, the PSD of the stellar fringes is computed numerically over the K band
using 10 nm spectral bins. We take here into account the total transmission of the interferometer
and the visibility of the fringes for each wavelength. The total power is then integrated
and gives a numerical broadband visibility function ${V_K}^2(B, \theta_{\rm UD})$ where $B$ is
the projected baseline, and $\theta_{\rm UD}$ the UD angular diameter.
To derive the $\alpha$\,Cen UD diameters, we make a
classical $\chi^2$ minimization between our $(B,{V_K}^2)$ measurements and the
${V_K}^2(B, \theta_{\rm UD})$ function while changing the value of $\theta_{\rm UD}$.

Fig.~\ref{FigAlfCenA_16m} shows the complete visibility
curve of the UD model fit to the $\alpha$\,Cen data, together with the primary calibrator $\theta$\,Cen.
The detail of the visibility curve of $\alpha$\,Cen\,A shown on Fig.~\ref{FigAlfCenA_66m}
demonstrates that the visibility  never goes down to zero for any baseline,
due to the bandwidth smearing effect. The minimum squared visibility is 0.15\%,
for a baseline length of aproximately 66.5m.
Fig.~\ref{FigAlfCenB_66m} shows an enlargement of the visibility points obtained on $\alpha$\,Cen\,B.
The final UD angular diameters for the two stars and the corresponding effective wavelengths
are given in Table~\ref{eff_lambda}.
\begin{table}
\caption[]{Uniform disk angular diameters of $\alpha$\,Cen\,A and B in the $K$ band
derived from the VINCI/VLTI observations.}
\label{eff_lambda}	\begin{tabular}{lccc}         &$\alpha$~Cen A &$\alpha$~Cen B\\ 
\hline \\
$\theta_{\rm UD}$ (mas) & $8.314 \pm 0.016$ & $5.856 \pm 0.027$ \\
\end{tabular}
\end{table}
\subsection{Limb darkened angular diameters}\label{LD_diams}
In this section, we describe two methods to compute the
LD angular diameter: through a conversion factor (classical approach), and through a
visibility fit taking directly the limb darkening into account.
\subsubsection{LD/UD conversion factor}\label{conversionLDUD}
The simplest approach to retrieve the limb darkened diameter
from an interferometric UD measurement goes through the computation of the
conversion factor $\rho$ defined by:
\begin{equation}
\rho = \frac{\theta_{\rm LD}}{\theta_{\rm UD}}
\end{equation}
$\rho$ is deduced from stellar atmosphere luminosity profiles that are computed using
radiative transfer modeling codes. These profiles are published in tables as a function of the
wavelength band (e.g. Claret \cite{claret00}).
One limitation of the description of the LD visibility curve of the star by a single parameter
is that it assumes that the visibility curve of the UD and LD models have the same intrinsic shape.
This is not exactly the case near and especially after the first minimum of the visibility function.
However, this approximation is satisfactory for compact stellar atmospheres
such as the ones of $\alpha$\,Cen stars. Hanbury Brown et al. (\cite{hanbury}) have shown
that the linear limb darkening coefficient $u$ can be translated into the conversion
factor $\rho$ through the approximate formula:
\begin{equation}
\rho = \sqrt{\frac{1-u/3}{1-7u/15}}
\end{equation}
These authors quote a maximum error of $\pm$0.2\% for this approximate formula, that is in general very
satisfactory, but for the particular case of $\alpha$\,Cen\,A, this uncertainty is comparable to
our final error bar on the UD diameter.
Different values of linear limb darkening conversion factors are given in Table~\ref{LD_conversion},
based on successive versions of the Claret et al. models (\cite{claret95}, \cite{claret98}, \cite{claret00}).
Except for the older Claret (\cite{claret95}) values, that do not take metallicity and turbulence
velocity into account, the two other results are very close to each other.
\begin{table}
      \caption[]{Linear LD/UD conversion factors for $\alpha$\,Centauri. The assumed physical
parameters to match Claret's (\cite{claret00}) grid are the closest ones
to those of Th\'evenin et al. (\cite{thevenin02}).}
	\label{LD_conversion}	\begin{tabular}{lcc}
	\hline  Model from Claret~(\cite{claret00})  &$\alpha$~Cen A &$\alpha$~Cen B\\ 
	\hline
	$\rm T_{\mathrm{eff}}$ (K) & 5750 & 5250 \\
	$\log(g)$ (cm.s$^{-2}$) & 4.5 & 4.5 \\
	$\log(M/H)$ (dex) & 0.2 & 0.2 \\
	$V_{T}$ (km/s) & 2.0 & 2.0 \\
	\hline
	\hline
	Reference & $\rho_A$ & $\rho_B$\\
	\hline
        Claret et al. (\cite{claret95}) & 102.047\% & 102.299\% \\
        Claret (\cite{claret98}) &102.388\%  & 102.723\% \\
        Claret (\cite{claret00}) & 102.355\% & 102.635\% \\
	\hline
        \end{tabular}
\end{table}

In order to account for a possible systematic error in the determination of the limb darkening
parameter, we allow a $\pm0.1$\% uncertainty to propagate into the computation
of the limb darkened diameter of $\alpha$\,Cen. It should be noted that the coefficients for
both stars originate from the same Kurucz's model atmosphere computations of Claret~(\cite{claret00}),
and are therefore likely to have a good intrinsic consistency.
\subsubsection{Limb darkened disk visibility fit}
\begin{figure}
   \centering
   \includegraphics[bb=0 0 360 144, width=8.5cm]{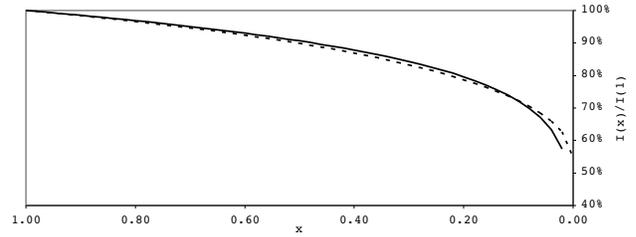}
      \caption{Intensity profile of $\alpha$\,Cen\,A from the four-parameters limb darkening
law of Claret (\cite{claret00}) (dashed line) and the corresponding $\alpha = 0.1417$ power law.}
         \label{LD_A}
   \end{figure}

Hestroffer (\cite{hestroffer}) has chosen another approach by computing the analytical
expression of the visibility function for a single-parameter power law intensity profile
$I_{\lambda}(\mu) = \mu^{\alpha}$ (with $\alpha \ge 0$) where $\mu = \cos(\theta)$
is the cosine of the azimuth of a surface element of the star.
This simplification allows this author to derive the analytical expression of the visibility
function corresponding to a power law limb darkened disk:
\begin{equation}
V_{\nu} (x) = \Gamma(\nu+1) \frac{J_{\nu}(x)}{(x/2)^{\nu}}
\end{equation}
where $x = \pi B \theta / \lambda$ is the spatial frequency, $\nu = \alpha/2 +1$,
and $J_{\nu} (x)$ is the Bessel function of the first kind. As the intensity profiles produced by
the most recent atmosphere models are close to power laws, as shown on Fig.~\ref{LD_A},
the power law fitting procedure gives good results.
Claret~(\cite{claret00}) has computed a four parameters law that aproximates very well the
most recent Kurucz models for observations in the K band.
Using this law gives a value of $\alpha = 0.1417$ for $\alpha$\,Cen\,A
and $\alpha = 0.1598$ for $\alpha$\,Cen\,B. 

The final precision on $\rho$ is better than with the previous linear approximation,
but as for the conversion coefficient approach presented in Sect.~\ref{conversionLDUD}, we
propagate an uncertainty of $\pm$ 0.1 \% to the final LD angular diameter to account for
a possible bias.

Practically, the fit is achieved on the calibrated visibilities listed in Tables
\ref{table_alfcenA} and \ref{table_alfcenB} by a classical  $\chi^{2}$ minimization procedure.
The product of this fit is directly the LD angular diameter of the star, without the
intermediate step of the uniform disk model.



%
%

\subsection{Rotational distortions}
As the VINCI/VLTI measurements have been obtained mostly at the same azimuth (roughly N-S),
a possible source of bias could be the presence of flattening on the stellar disks due to rotational
distortion. The estimated equatorial rotation periods for $\alpha$~Cen A
and B are 22 and 41 days respectively (Morel et al. \cite{mpl00}), bracketing the solar value.
The corresponding small rotational velocities rule out any flattening at a significant level,
and therefore no correction has been applied to our measurements.

\subsection{Summary of angular diameter values}
Table \ref{LD_summary} gives the limb darkened angular diameters derived from the LD/UD
conversion factors and from the analytical LD visibility function (Hestroffer \cite{hestroffer}).
This last method is assumed in the following sections. All values take the 
bandwidth smearing effect into account.
\begin{table}
      \caption[]{Summary of limb-darkened angular diameters from different computation methods. 
Both methods are based on the Claret (\cite{claret00}) coefficients.
The fitting results using the analytical Hestroffer (\cite{hestroffer}) formula are assumed
in the rest of this paper.}
	\label{LD_summary}	\begin{tabular}{lc}
	\hline         LD computation method & $\alpha$~Cen A (mas) \\ 
	\hline
        Hanbury Brown et al. (\cite{hanbury}) & 8.517 $\pm$ 0.021 \\
        {\bf Hestroffer (\cite{hestroffer}) analytical} & {\bf 8.511 $\pm$ 0.020} \\
	\hline
         &$\alpha$~Cen B (mas) \\ 
	\hline
        Hanbury Brown et al. (\cite{hanbury}) & 6.010 $\pm$ 0.030 \\      
        {\bf Hestroffer (\cite{hestroffer}) analytical} & {\bf 6.001 $\pm$ 0.034} \\
	\hline
       \end{tabular}
\end{table}
%
\section{Comparison of asteroseismic and interferometric linear diameters}
\subsection{Parallax from the literature}
To convert the angular diameter into a linear value, it is necessary to know
the parallax of the star. The $\alpha$\,Cen system being very nearby
($D = 1.3$ pc), the precision on the measurement of 
its trigonometric parallax is potentially very good.
Unfortunately, some discrepancies have appeared between the
most recently published values (Table~\ref{alfCen_parallaxes}).
In particular, the original {\it Hipparcos} parallax (Perryman et al. 1997) and
the value by Pourbaix et al. (\cite{pourbaix99}) are significantly different
from the reprocessing of the {\it Hipparcos} data
by S\"oderhjelm (\cite{soderhjelm}), by more than 3$\sigma$.
A difficulty with the {\it Hipparcos} satellite measurement is due to the large brightness of 
the $\alpha$\,Cen pair. The light from one of the stars possibly contaminated the measurement 
on the other, leading to a systematic bias that may not have been propagated
properly to the final error bars. In Sect.~\ref{plx_VLTI}, we adopt the parallax value
from S\"oderhjelm (\cite{soderhjelm}), who took this effect into account.
\begin{table}
      \caption[]{Parallax values of $\alpha$\,Cen from the litterature.}
	\label{alfCen_parallaxes}	\begin{tabular}{cl}
\hline
Value (mas) & Author\\
\hline
750 & Heintz (\cite{heintz58}, \cite{heintz82})\\
749 $\pm$ 5 &  Kamper \& Wesselink (\cite{kamper78})\\
750.6 $\pm$ 4.6 & Demarque et al. (\cite{demarque86})\\
742.1 $\pm$ 1.4 & Perryman et al. (\cite{hip})\\
737.0 $\pm$ 2.6 & Pourbaix et al. (\cite{pourbaix99}) \\
747.1 $\pm$ 1.2 & S\"oderhjelm (\cite{soderhjelm})\\
\hline
\end{tabular}
\end{table}

As a remark, the semi-major axis of the orbit of the two stars
$a = 17.59 \pm 0.03$ AU (Pourbaix et al. \cite{pourbaix99}) is totally negligible compared 
to the distance $D$ to the couple ($a / D = 0.006\%$), 
therefore the two stars can be considered at the same distance.
\subsection{Linear diameters}\label{plx_VLTI}
Considering the parallax of $747.1 \pm 1.2$ mas from S\"oderhjelm (\cite{soderhjelm}),
it is now possible to compute the linear diameters of $\alpha$\,Cen\,A and B (in solar units)
from the two LD angular diameters determined interferometrically. They
are found to be:
\begin{equation}
D[A] = 1.224 \pm 0.003 \ D_{\odot}
\end{equation}
\begin{equation}
D[B] = 0.863 \pm 0.005 \ D_{\odot}
\end{equation}
and can be compared to the linear diameters proposed by Th\'evenin et al.~(\cite{thevenin02}):
\begin{equation}
D[A] = 1.230 \pm 0.003 \ D_{\odot}
\end{equation}
\begin{equation}
D[B] = 0.857 \pm 0.007 \ D_{\odot}
\end{equation}

The theoretical diameters are only +1.3$\sigma$ and -0.7$\sigma$ away from the observed values.
Both interferometric diameters and those deduced from the photometric calibration
constrained by asteroseismic frequencies therefore agree very well. 

These two model diameters are derived using the CESAM code, and are defined as the radii at which
the temperature in the atmosphere is equal to the effective temperature of the star. Computing
the layer where the continuum at 2.2~$\mu$m is formed leads to temperatures
close to $T_{\mathrm{eff}}$, therefore the CESAM diameters can be directly compared
to those measured by the VLTI at 2.2~$\mu$m.

\subsection{Self consistent parallax}
From our angular diameter measurements and the asteroseismic
diameter estimations, we can also derive directly the parallax of the couple. 
The simple formula linking the limb darkened angular diameter $\theta_{\rm {LD}}$ (in mas),
the linear diameter $D$ (in $D_{\odot}$) and the parallax $\pi$ (in mas) is:
\begin{equation}\label{distance_eq}
\theta_{\rm {LD}} = 9.305.10^{-3} \ D \ \pi
\end{equation}
A least squares fit is computed between the LD angular diameters
from the VLTI and the linear diameters from Th\'evenin et al.~(\cite{thevenin02}).
We find an optimal parallax of $745.3 \pm 2.5$ mas,
that differs slightly from the original {\it Hipparcos} value by +1.1$\sigma$ (Perryman et al. \cite{hip}),
from Pourbaix et al. (\cite{pourbaix99}) by +2.3$\sigma$, and from S\"oderhjelm (\cite{soderhjelm})
by only -0.7$\sigma$. The resulting values of angular and linear diameters
are given in Table~\ref{alfCen_parallaxes_VLTI}. The difference between theoretical and linear
diameters for the self-consistent parallax is limited to +0.5$\sigma$ and -0.9$\sigma$,
respectively for $\alpha$\,Cen~A and B.
\begin{table}
      \caption[]{Parallax of $\alpha$~Cen from VINCI/VLTI and asteroseismological observations,
and the corresponding self-consistent stellar parameters. Linear diameters are taken from the
asteroseismology study by Th\'evenin et al.~(\cite{thevenin02}).}
	\label{alfCen_parallaxes_VLTI}	\begin{tabular}{lcc}
\hline
 Derived parallax & $745.3 \pm 2.5$ mas \\\hline
	 & $\alpha$\,Cen\,A & $\alpha$\,Cen\,B \\ 
\hline
VINCI LD size (mas) & $8.511 \pm 0.020$ & $6.001 \pm 0.034$\\
Model LD size (mas) & $8.530 \pm 0.035$ & $5.943 \pm 0.052$\\
\hline
VINCI diameter ($D_{\odot}$) & $1.227 \pm 0.005$ & $0.865 \pm 0.006$\\
Model diameter ($D_{\odot}$) & $1.230 \pm 0.003$ & $0.857 \pm 0.007$\\
\hline \\
\end{tabular}
\end{table}
\subsection{Ratio of  $\alpha$\,Cen\,A and B radii}
Contrary to the linear diameters themselves, their ratio is independent of the actual
parallax of the system. Therefore, part of the systematic uncertainties can
be removed by using this observable as a comparison basis between
the observations and the models.
For $\alpha$\,Cen\,A and B, we have access to a very good quality parallax,
and the uncertainty introduced there is relatively small.
On the other hand, when measuring a farther double
or multiple star, the parallax may be unknown, or known only with a bad precision. In this
case, comparing the ratio of the stellar diameters will give much stronger constraints
to the stellar structure models than the individual values. This technique
is also applicable to the interferometric measurement of stars in clusters, within
which the distance can be assumed to be uniform.
From the limb-darkened values listed in Table \ref{LD_summary},
we obtain the following ratio between the angular diameters of $\alpha$\,Cen\,A and B:
\begin{equation}
\frac{\theta_{\rm LD}[A]}{\theta_{\rm LD}[B]} = 1.418 \pm 0.009
\end{equation}
This value can be compared to the ratio of linear radii from the
Th\'evenin et al. (\cite{thevenin02}) models that is:
\begin{equation}
\frac{R[A]}{R[B]} = 1.435 \pm 0.014
\end{equation}
We therefore find a slight diameter excess of $\alpha$\,Cen B at a level of $1.0\sigma$.
%
\subsection{Masses and evolutionary models}\label{sec:res}
As emphasized by Th\'evenin et al. (\cite{thevenin02}),
the seismological analysis gives strong constraints on masses and
on the age of the system when combined with spectro-photometric measurements.
To achieve this, one derives from the set of oscillation frequencies, 
one ``large'' and two ``small'' frequency spacings.
The large frequency spacing is a difference
between frequencies of modes with consecutive radial order $n$:
$\Delta\nu_\ell(n) \equiv \nu_{n, \ell} - \nu_{n-1,\ell}.$ 
In the high frequency range, i.e. large radial orders, 
$\Delta\nu_\ell$ is almost constant with  a mean value $\Delta \nu_{0}$, strongly
related to the mean density of the star, i.e. to the mass and the radius.
The small separations are very sensitive to the physical conditions in the core of the star
and consequently to its age.
These frequencies measured  for the star A
have largely forced the spectro-photometric calibration to decrease
the masses of the stellar system $\alpha$\,Cen, leading to the following values:
$M_{\rm A}=1.100\pm0.006\,M_\odot$
and $M_{\rm B}=0.907\pm0.006\,M_\odot$ (Th\'evenin et al. \cite{thevenin02})
close to those adopted by Guenther \& Demarque~(\cite{gd00}) and Kim~(\cite{gd00}).
The mass of the B component departs significantly by 3\% from the value published by
Pourbaix et al.~(\cite{pnm02}).
   \begin{figure}
   \centering
    \includegraphics[width=6.0cm, angle=-90]{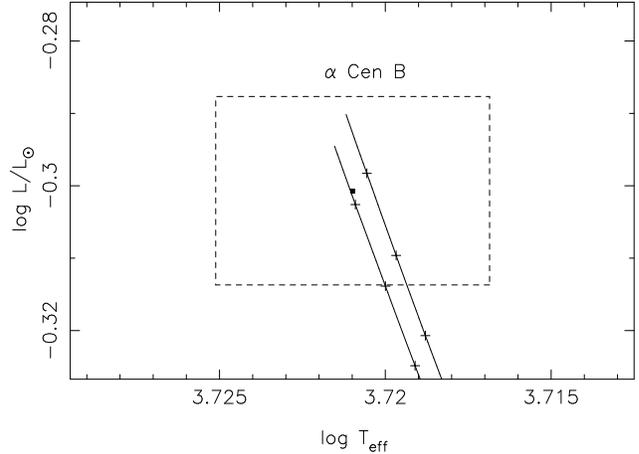}
      \caption{HR diagram of $\alpha$\,Cen\,B. The line on the right corresponds to 
a mixing length of $\lambda=0.96$ and a mass of 0.909~M$_{\odot}$, the line on the left 
corresponds to the values published in Th\'evenin et al. \cite{thevenin02} ($\lambda=0.98$, 0.907~M$_{\odot}$).}
         \label{HR_fig}
   \end{figure}

Using the orbital properties of the binary and also spectro-velocimetric curves,
Pourbaix et al. (2002) have derived the masses of each components
($M_{\rm A}=1.105\pm0.007\,M_\odot$, $M_{\rm B}=0.934\pm0.006\,M_\odot$).
We note that Thoul et al. (\cite{thoul03}) have recently proposed a model of the
binary system using these masses and spectro-photometric constraints different from that of
Th\'evenin et al.~(\cite{thevenin02}). They were able to reproduce the seismic frequencies
of $\alpha$\,Cen\,A, but the model they propose does not take into account the
helium and heavy elements diffusion.

Because the interferometric diameter of $\alpha$~Cen B is a little larger than those
deduced from the CESAM model, we explored the possibility to decrease this difference
by changing the mixing length of the B model from $\lambda$=0.98 to $\lambda$=0.96,
and by increasing the mass of the star from 0.907 to 0.909~M$_{\odot}$. 
These modifications do not change the calibration of $\alpha$\,Cen\,A.
We took care in this process to keep the star B in its error box on the
HR diagram (Fig.~\ref{HR_fig}). It results from this new mass a diameter
that is closer to the interferometric one: 
0.863 $D_{\odot}$ or $5.999 \pm 0.050$ mas (parallax from S\"oderhjelm~\cite{soderhjelm}).
The effective temperature is found to be 5262 K, identical to the adopted
spectroscopic one $\rm T_{\mathrm{eff}} = 5260$ K.
Our results confirm that the mass of the B component is probably close to 0.907 M$_{\odot}$,
as reported by Th\'evenin et al.~(\cite{thevenin02}).
A similar exercice is not possible if we adopt the mass of 0.934 M$_{\odot}$ derived
by Pourbaix et al.~(\cite{pnm02}).
%
\section{Conclusion \label{conclusion}}
We have determined the angular diameters of $\alpha$\,Cen A and B using the VINCI/VLTI
instrument, to a relative precision of 0.2\% and 0.6\%, respectively.
The low values of the $\alpha$\,Cen\,A visibilities allowed us to match our statistical visibility error
to the calibration uncertainty.
This is an optimal situation for the angular diameter
measurement, that would not have been feasible with a higher visibility.
Calibrating with a fainter and smaller unresolved star would also not have been
efficient, as we would have degraded significantly our statistical precision.
There is still a compromise, as the low visibilities of $\alpha$\,Cen\,A imply a
slightly degraded statistical precision, but E0-G1 has proved to be
a well suited baseline for the simultaneous measurement of the angular
diameters of $\alpha$\,Cen\,A and B.

The comparison of these interferometric diameters with the values derived using asteroseismic
constraints shows a good agreement when adopting the parallax determined by
S\"oderhjelm (\cite{soderhjelm}). In particular, our diameters are compatible with
the masses proposed by Th\'evenin et al.~(\cite{thevenin02}) for both stars.
In the near future, asteroseismic observations of the large frequencies spacing
$\Delta\nu_\ell$ of $\alpha$\,Cen\,B will complete the calibration of this binary system.
Simultaneously, the very long baselines of the VLTI (up to 200m) will allow us to
measure directly the limb darkening of these two stars, and therefore derive
the photospheric diameter without using a stellar atmosphere model.
This work demonstrates that the combination of the interferometry and asteroseismology
techniques can provide strong constraints on stellar masses and
other fundamental parameters of stars.
\begin{acknowledgements}
We are grateful to V. Coud\'e du Foresto and G. Perrin for fruitful discussions regarding
the analysis of the VINCI data, and to M. Wittkowski for his useful comments on the
limb darkening question.
We thank also the ESO VLTI team for operating the VLTI, and for making the
commissioning data publicly available.

The interferometric measurements have been obtained using the Very Large Telescope
Interferometer, operated by the European Southern Observatory at Cerro Paranal, Chile.
The VINCI public commissioning data used in this paper
has been retrieved from the ESO/ST-ECF Archive (Garching, Germany).
This research has also made use of the SIMBAD database at CDS, Strasbourg (France) and
of the WDS database at USNO, Washington, DC (USA).
\end{acknowledgements}

%
\end{document}